\documentclass[aps,pra,twocolumn,superscriptaddress,showpacs]{revtex4-1}
\usepackage{graphicx,amsmath,amssymb,ulem,color,braket}
\begin{document}

\title{Two Copies of the Einstein-Podolsky-Rosen State of Light\\ Lead to Refutation of EPR Ideas}

\author{Krzysztof Roso\l{}ek}
\address{Institute of Theoretical Physics and Astrophysics, University of Gda\'nsk, ul.~Wita Stwosza 57, 80-952 Gda\'nsk, Poland}
\author{Magdalena Stobi\'nska}
\email{magdalena.stobinska@gmail.com}
\thanks{Corresponding author}
\address{Institute of Theoretical Physics and Astrophysics, University of Gda\'nsk, ul.~Wita Stwosza 57, 80-952 Gda\'nsk, Poland}
\address{Institute of Physics, Polish Academy of Sciences, Al.~Lotnik\'ow 32/46, 02-668 Warsaw, Poland}
\author{Marcin Wie\'sniak}
\address{Institute of Theoretical Physics and Astrophysics, University of Gda\'nsk, ul.~Wita Stwosza 57, 80-952 Gda\'nsk, Poland}
\author{Marek \.Zukowski}
\address{Institute of Theoretical Physics and Astrophysics, University of Gda\'nsk, ul.~Wita Stwosza 57, 80-952 Gda\'nsk, Poland}

\begin{abstract}
Bell's theorem applies to the normalizable approximations of the original Einstein-Podolsky-Rosen (EPR) state. The constructions of the proof require measurements difficult to perform, and dichotomic observables. By noticing the fact that the four mode squeezed vacuum state produced in type II down-conversion can be seen both as two copies of approximate EPR states, and also as a kind of polarization supersinglet, we show a straightforward way to test violations of the EPR concepts with direct use of their state. The observables involved are simply photon numbers at outputs of polarizing beam splitters. Suitable chained Bell inequalities are based on the geometric concept of distance. For a few settings they are potentially a new tool for quantum information applications, involving observables of a nondichotomic nature, and thus of higher informational capacity. In the limit of infinitely many settings we get a Greenberger-Horne-Zeilinger-type contradiction: EPR reasoning points to a correlation, while quantum prediction is an anticorrelation. Violations of the inequalities are fully resistant to multipair emissions in Bell experiments using parametric down-conversion sources.
\end{abstract}

\maketitle

{\it Introduction.} 
Quantum phenomena are counterintuitive and the formalism is even more. Predictions of quantum mechanics (QM) are of statistical nature: there is no deterministic theory of response of individual systems to all possible experiments. Some quantum predictions seem paradoxical~\cite{QP}. 

The Einstein-Podolsky-Rosen paradox~\cite{EPR} was an attempt to show that the quantum description of reality cannot be complete. Elements of reality, properties of a system, which can be established with perfect accuracy without in any way disturbing it, were suggested to be the missing component of the theory. EPR used perfectly correlated systems to argue that such elements are derivable from quantum predictions and the principle of relativistic locality. There were some additional tacit assumptions in the reasoning of EPR, like the freedom of the experimentalist to choose the observable to be measured, and the equivalence of the actual experimental situation realized for the given individual system, and a complementary one \cite{CASLAV-ZUK}. The second of these was challenged  by Bohr \cite{BOHR}: ``... there is essentially the question of an influence on the very conditions which define the possible types of predictions regarding the future behavior of the system... In fact, it is the mutual exclusion of any two experimental procedures, permitting unambiguous definition of complementary physical quantities, which provides room for new physical laws the coexistence of which at first sight appear irreconcilable with the basic principles of science.''

50 years ago, Bell showed  a technical flaw in the EPR reasoning~\cite{Bell}: in Bohm's version of the paradox \cite{BOHM}, for a two-spin $1/2$ singlet,  elements of reality are  incompatible with QM. They must satisfy Bell's inequalities, while quantum predictions violate them. A more striking contradiction is  by the Greenberger, Horne and Zeilinger (GHZ)~\cite{GHZ, MERMIN}: for three spin-$1/2$ particles,  if elements of reality exist,
 then $1=-1$. These results led to an `industry' which uses violations of Bell inequalities in practical applications: e.g. reduction of communication complexity~\cite{Zuk}, randomness generation~\cite{Rand}, {\em device-independent} quantum cryptography~\cite{QKD}, and as entanglement `witnesses' ~\cite{HOR-RMP,PAN}.

A question remained unresolved for many years: Does the Bell's theorem hold true also for the EPR state? The momentum representation of it is $\Psi(p_1,p_2)=\delta(p_1+p_2)$, where $p_i$ is the momentum of $i$-th `particle'. Such singular objects do not exist in the Hilbert space. Nevertheless, they can be approximated by well-behaved functions, which in some limit give  $\delta(p_1+p_2)$. In \cite{BELL2} Bell shows that the Wigner distribution for the EPR state is non-negative in the entire phase space, thus there is no chance for a Bell inequality violation, as we have explicit local hidden variable model. 

Meanwhile, Reid and Drummond~\cite{REID,REID2} showed that the state emitted by a non-degenerate optical parametric amplifier, {\it two mode} squeezed vacuum, is an optical approximation of the EPR state. This opened prospects for  observing approximate `original' EPR correlations. 

Bell's theorem for approximate EPR states was finally given in \cite{COHEN} and \cite{BANASZEK}. The idea was to use different observables than the ones discussed by EPR. Cohen \cite{COHEN} used an approach which requires a highly specific interferometer, or coupling of the EPR state to a pair of spin $1/2$ ancillas. In \cite{BANASZEK} observables with singular Wigner representations were used (parity operators, or
no count events, highly dependent on losses). In both cases displacement was involved. It requires auxiliary coherent states,  and thus necessary filtering to get indistinguishability of photons from different sources, which introduces losses~\cite{LAIHO,BARTLEY}.

Below we review and reveal properties of the four-mode squeezed vacuum state (SV) related with EPR paradox. This leads us to formulation of specific chained Bell inequalities, violated by the SV state.
The non-classical phenomena related with SV can be used in quantum information and communication, and allow for a GHZ-like argument. The SV can be interpreted both as approximate two copies of the EPR state or a polarization super-singlet. We conclude with a discussion and interpretation of our results, and remarks on feasibility of their experimental demonstration. We emphasize that we do not aim at seeking 
robust phenomena leading to loophole-free Bell tests,
but rather to reveal  exciting phenomena linked with the four-mode SV state. It constitutes both a realistic resource for quantum technologies, and can lead to exciting case studies in foundations of quantum theory.

{\it Four-mode SV singlet state.}
 The standard method of its generation employs a type II parametric down-conversion (PDC)  in a nonlinear crystal pumped by a laser beam~\cite{PAN}. This process is described by the Hamiltonian $\mathcal{H} =  i g (a_H^{\dagger} b_V^{\dagger} +e^{i\phi} a_V^{\dagger} b_H^{\dagger}) + \mathrm{H.c.}$, where in the notation for creation operators letters $a,b$ stand for distinct spatial beams, and subscripts $H,V$ for linear polarizations; the coupling $g$ is proportional to the pumping field. We assume  $e^{i\phi}=-1$. The output state is a superposition of maximally entangled $2N$-photon polarization singlet states
\begin{align}
\lvert\Psi^{(-)}\rangle ={}&  \sum_{N=0}^{\infty} \lambda_N \lvert\psi^{(-)}_{N}\rangle,
\label{BSV}
\end{align}
where $\lambda_N= \cosh^{-2}\Gamma\sqrt{N+1} \tanh^{N}\Gamma$, $\sum_{N=0}^{\infty} \lambda_N^2 = 1$, 
\begin{eqnarray}
&\lvert\psi^{(-)}_{N}\rangle= \tfrac{1}{\sqrt{N+1}N!} (a^{\dagger}_Hb^{\dagger}_V-a^{\dagger}_V b^{\dagger}_H)^N\lvert0\rangle &\label{singletN} \\
& = \tfrac{1}{\sqrt{N+1}} \sum_{n=0}^N (-1)^n \lvert n_H,(N-n)_V\rangle_a \lvert (N-n)_H,n_V\rangle_b.&\nonumber\\ \nonumber
\end{eqnarray}
The symbol $\lvert n_H,(N-n)_V\rangle_a$ denotes $n$ horizontally and $N-n$ vertically polarized photons in beam $a$, similarly for   beam $b$. The  states $\lvert \psi^{(-)}_{N}\rangle$ contain $N$ photons in each beam. Polarization of each beam  is undefined. However,  due to equal photon numbers in the orthogonal polarizations beams are anti-correlated. The effective strength of the interaction is  $\Gamma=gt$,  where $t$ is the interaction time.

The unitary transformation generating $\Psi^{(-)}$ is given by  $e^{i\mathcal{H}t}$, and can be factorized as  $e^{i\mathcal{H}_{H,V}t}  e^{i\mathcal{H}_{V,H}t} $, where $\mathcal{H}_{H,V}=   i g (a_H^{\dagger} b_V^{\dagger}) + \mathrm{H.c.}$ and  $\mathcal{H}_{V,H}=   -i g (a_V^{\dagger} b_H^{\dagger}) + \mathrm{H.c.}$  The initial state is vacuum.  We get two  approximate EPR states, two squeezed two-mode vacua: one for modes $a_H$ and $b_V$ and the second one, for  $a_V$ and $b_H$.

\begin{figure}
\centering 
\includegraphics[height=3.2cm]{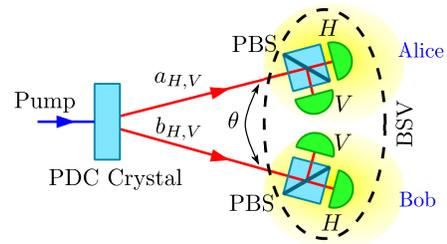}
\caption{Test of inequality (\ref{ChainBellIneq}) with four-mode squeezed vacuum state. PDC -- parametric down conversion crystal, PBS -- polarizing beam splitter. The detectors measure photon numbers.}
\label{fig:BSV-setup}
\end{figure}

{\it EPR elements of reality vs. $\lvert \Psi^{(-)}\rangle$.} 
Consider a Bell experiment in Fig.~\ref{fig:BSV-setup}. Two spatially separated observers, Alice and Bob observe radiation of a pulse pumped source producing the SV state. They control the orientation of their local polarizing beam splitters, $\theta_{A}$ and $\theta _{B}$, respectively, and count photons at their outputs. The result of the local measurement for run $k$ is a certain number of $\theta_{A}$ linearly polarized photons counted at Alice's side $n^{(k)}(\theta_A)$ and at Bob's side $m^{(k)}(\theta_B)$. Since Hamiltonian is invariant with respect to the choice of  pairs of orthogonal (generally elliptic) polarizations:
 $\mathcal{H}=ig(a^{\dagger}_\theta b^{\dagger}_{\theta^\perp}-a^{\dagger}_{\theta^\perp}b^{\dagger}_\theta) + \mathrm{H.c.}$, where  $\theta^\perp=\theta+\frac{\pi}{2},$ 
 if $\theta_B=\theta_A+\frac{\pi}{2}$ then $n^{(k)}(\theta_A)=m^{(k)}(\theta_A+\frac{\pi}{2})$.  In the above notation $\theta=0$ denotes horizontal polarization $H$, etc. Recall that the  two-photon polarization singlet state of  Bohm  \cite{BOHM},  $\frac{1}{\sqrt{2}}(a^{\dagger}_Hb^{\dagger}_V-a^{\dagger}_Hb^{\dagger}_V)\lvert 0\rangle$, is invariant with respect to $U\otimes U$  polarization rotations. The four mode SV posses the same invariance. Thus, it is a kind of polarization super-singlet,  with undefined number of photons.

This feature of $\lvert\Psi^{(-)}\rangle$ allows for an EPR-like reasoning with different observables than the ones considered in earlier works.  A distant  measurement at Alice's side with setting $\theta_A$ can fix Bob's value for the $k$-th run for his setting $\theta_B=\theta_A+\frac{\pi}{2}$, without measuring it, and {\it vice versa}. Here, we use the property $n^{(k)}(\theta_A)=m^{(k)}(\theta_A+\frac{\pi}{2})$. 
This suggest that $n^{(k)}(\theta_A)$ and $m^{(k)}(\theta_B)$ are {\it elements of reality}. They seem to exist for  any $\theta_A$ and  $\theta_B$.

This EPR-like reasoning is inconsistent. A Bell inequality is satisfied by the elements of reality, and violated by quantum predictions.  The double EPR-like-supersinglet $\lvert\Psi^{(-)}\rangle$ leads to predictions which disagree with the ideas of EPR. 

{\it Chained Bell inequalities.} 
The inequalities  are based on the concept of distance. Any properly defined distance satisfies polygon inequalities. Take two stochastic variables $V(\lambda)$ and $W(\lambda)$, governed by a joint probability  $\rho(\lambda)$. Their `separation' can be measured by $D(V,W)=\int \lvert V(\lambda)-W(\lambda)\rvert\rho(\lambda)d\lambda $. This function satisfies all defining properties of a distance: $D(V,V)=0$, $D(V,W)=D(W,V)\geq0$  and  the triangle inequality $D(V, Z)\leq D(V,W)+D(W,Z)$. The last property is due to the fact that for any three  numbers $a,b,c$ one has: $\lvert a-c\rvert\leq \lvert a-b\rvert+\lvert b-c\rvert$. 

\begin{figure}
\center
\includegraphics[height=1.7cm]{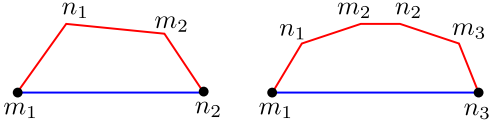}
\caption{Polygon inequalities for distance. The sum of the lengths of the red segments is greater than the length of the blue segment.}
\label{fig:polygon}
\end{figure}

Alice and Bob choose freely between several local settings of their polarizers, $\theta_{A_i}$ and $\theta_{B_j}$, respectively. For a concise notation, we denote the elements of reality associated with  the $k$-th run of the experiment by $n_i^{(k)}=n^{(k)}(\theta_{A_i})$ and $n_j^{(k)}=n^{(k)}(\theta_{B_j})$.

The triangle inequality implies polygon inequalities, illustrated in Fig.~\ref{fig:polygon}. Let $i,j=1,...,L$. A polygon inequality for numbers representing the elements of reality, takes the form
\begin{align}
\sum_{i=1}^{L} \lvert m_{i}^{(k)}\!\!- n_{i}^{(k)}\rvert + \sum_{i=1}^{L-1} \lvert m_{i+1}^{(k)}\!\!- n_{i}^{(k)}\rvert \geq \lvert m_{1}^{(k)}\!\!-n_{L}^{(k)}\rvert.
\label{ChainBellIneq1}
\end{align}
For averages, $\langle \lvert m_{i} - n_{j}\rvert \rangle=\frac{1}{R}\sum_{k=1}^R\lvert m_{i}^{(k)} - n_{j}^{(k)}\rvert $, where $R$ is the number of runs, we get:
\begin{align}
\sum_{i=1}^{L}\langle \lvert m_{i}- n_{i}\rvert\rangle + \sum_{i=1}^{L-1}\langle \lvert m_{i+1}- n_{i}\rvert \rangle \geq \langle\lvert m_{1}-n_{L}\rvert\rangle.
\label{ChainBellIneq}
\end{align}
Thus we arrive at distance-based Bell inequalities (for different chained inequalities see \cite{BC}). 

Inequality (\ref{ChainBellIneq})  also holds for  local hidden variable (LHV) approaches.  If variables  $m_{i}$ and $n_{j}$ depend on  some hidden parameters $\lambda$, and  their `distance' equals
\begin{equation}
\langle \lvert m_{i}-n_{j} \rvert \rangle = \int d\lambda \rho_{hv} (\lambda )\big\lvert m_{i}(\lambda )-n_{j}(\lambda ) \big\rvert,
\label{aver_mod}
\end{equation}
where $\rho_{hv} (\lambda )$ is a probability distribution.

Within quantum theory, in (\ref{ChainBellIneq}) we shall use as observables photon number operators $a_i^{\dagger}a_i$ (Alice) and $b_j^{\dagger}b_j$ (Bob). The measurement settings by Alice and Bob will be defined by  $a_i = \cos \theta_{{A}_{i}} \, a_H + \sin \theta_{{A}_{i}} \, a_V$, 
and $b_i= -\sin \theta_{{B}_{i}} \, b_H + \cos \theta_{{B}_{i}} \, b_V$. The inequality (\ref{ChainBellIneq}) requires the following holds
\begin{align}
\mathrm{LHS} =&\sum_{i=1}^{L}\langle \lvert a^{\dagger}_{i}a_{i} - b^{\dagger}_{i}b_{i}\rvert \rangle + \sum_{i=1}^{L-1} \langle \lvert a^{\dagger}_{i+1}a_{i+1} - b^{\dagger}_{i}b_{i}\rvert \rangle \nonumber\\
&\geq \langle \lvert a^{\dagger}_{1}a_{1} - b^{\dagger}_{L}b_{L}\rvert \rangle = \mathrm{RHS}.
\label{eq:ineq}
\end{align}

\begin{figure}
\centering
\includegraphics[height=3cm]{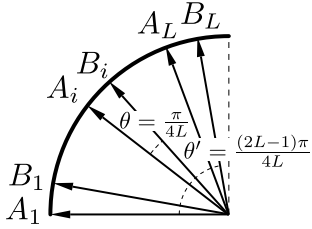}
\caption{Measurement settings for Alice (A) and Bob (B) for testing the distance Bell inequality (\ref{eq:ineq}).}
\label{fig:kolo}
\end{figure}

{\it Violations of   (\ref{ChainBellIneq}) by the supersinglet $\Psi^{(-)}$.} 
The measurements  for Alice and Bob are displayed in Fig.~\ref{fig:kolo}. We set $\theta_{A_1}=0$ and $\theta_{B_1}=\theta=\tfrac{\pi}{4L}$ The relative angle between the polarization settings by Alice $\theta_{A_i}$ and Bob $\theta_{B_i}$ we put as constant, equal to  $\theta$. Each subsequent setting of Alice and Bob changes by $2\theta$. Thus, the angle between $\theta_{A_{i+1}}$ and $\theta_{B_i}$ is also $\theta$. The angle between the first Alice's setting $\theta_{A_1}$ and the last of Bob's setting $\theta_{B_L}$ is set  to $\theta'=\tfrac{(2L-1)\pi}{4L}$. Due to  the $U\otimes U$ invariance of  $\Psi^{(-)}$, the quantum predictions for counts in $a$ and $b$ depend only on the relative angle, $\theta$  or $\theta'$. Note that,  for $\theta=0$,  perfect correlations (\ref{singletN}) between the orthogonal polarizations in beams $a$ and $b$ are observed. 

In the lossless case, Alice and Bob always  measure altogether, in the two outputs of local polarizers, $N$ photons each (we shall analyze losses later). For the settings $\theta_{A_1}=0$ and $\theta_{B_1}=\theta$, the probability $p_Q^N(n, m\mid \theta)$ to register $n$ photons in Alice's channel $H$ and $m$ in Bob's channel $\theta_{B_1}^{\perp}$, denoted below as ${V+\theta}$, reads
\begin{equation}
\Big\lvert\langle \psi^{(-)}_{N}\rvert\big(\lvert n_H, (N-n)_V\rangle_a\lvert (N-m)_{H+\theta }, m_{V+\theta}\rangle_b\big)\Big\rvert^{2}.
\label{FORMULA}
\end{equation}
As the components $|\psi^{(-)}_{N}\rangle$ do not mix up, 
we can consider (\ref{eq:ineq}) for each component separately, as effectively we have:
\begin{align}
\label{INEQ}
\mathrm{LHS} ={}&\sum_{n,m=0}^{N}\lvert m - n\rvert (2L-1) p_Q^N(n, m\mid \theta )\\  \geq&{}\sum_{n,m=0}^{N}\lvert m - n\rvert p_Q^N(n, m\mid \theta' )= \mathrm{RHS}.\nonumber
\end{align}

Let us estimate the RHS of (\ref{INEQ}) for a large number of settings $L$ (a long chain). Then, $\theta' \approx \frac{\pi }{2}$. We have $\frac{\pi }{2}$  in the limit $L \to \infty$, and Bob's $H$ is now $V$. Perfect anti-correlation is observed, $m=N-n$; one has $p_Q^N(n, N-n\mid \frac{\pi }{2}) =\tfrac{1}{N+1}$. Taking into account the summation over $n$ and $m$, the RHS grows linearly with~$N$.

To estimate the LHS of (\ref{INEQ}), notice that $\lvert n_H, (N-n)_V\rangle_a\lvert (N-m)_{H+\theta },m_{V+\theta}\rangle_b$ is proportional to ${a^{\dagger n}_H}{a^{\dagger(N-n)}_V} { b^{\dagger (N-m)}_{H+\theta }}{b^{\dagger m }_{V+\theta}}\lvert 0\rangle$. Since $ { b^{\dagger}_{H+\theta }}= b^{\dagger}_H\cos{\theta}+  b^{\dagger}_V\sin{\theta}$ and $ { b^{\dagger}_{V+\theta }}= b^{\dagger}_V\cos{\theta}-  b^{\dagger}_H\sin{\theta}$, for $\theta=0$ the perfect singlet correlations are recovered: $p_Q^N(n, m\mid \theta= 0)$ is non-zero only for $n=m$, and the average of $\lvert m-n\rvert$ vanishes. For $\theta\not=0$, all `new' terms in (\ref{FORMULA}) are proportional to {\it even} powers of $\sin{\theta}$. The `old' term proportional to $\cos^{2N}{\theta}$ does not contribute to $p^N_Q(n\neq m\mid\theta)$. Thus, the difference between $p^N_Q(n\neq m\mid0)=0$ and $p^N_Q(n\neq m\mid\theta)=0$ is a polynomial in $\sin{\theta}$ with the lowest power equal  to 2. As $\theta=\tfrac{\pi}{4L}$, the lowest order terms in the LHS of (\ref{INEQ}) behave as $(2L-1)\sin^2{\frac{\pi}{4L}}$ and tend to zero for large $L$. Higher order terms vanish even quicker. Therefore, the LHS approaches zero and, in the limit  $L \to \infty$,  we have an "all-versus-nothing" conflict with the prediction for the RHS.

\begin{figure}
\centering 
\includegraphics[height=3cm]{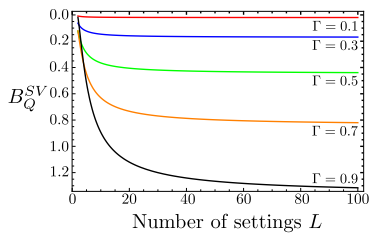}
\caption{The Bell parameter $B_{Q}$ as a function of number of settings $L$ evaluated for the entangled squeezed vacuum state. For $\Gamma=0.8$ the mean number of photons equals $1.6$.}
\label{LimitBSV}
\end{figure}

We may define a Bell parameter for the SV state as follows $B_{Q} = \sum_{N=0}^{\infty} \lambda_N^2 B_{Q}^N$, where $B_{Q}^N=\mathrm{LHS} - \mathrm{RHS}$ is computed for $|\psi_N^{(-)}\rangle$ state. For $L\rightarrow \infty$ and an odd $N$, $B_{Q}^N = -\frac{\frac{1}{2}N^{2}+N+\frac{1}{2}}{N+1}$ and for even, $B_{Q}^N = -\frac{\frac{1}{2}N^{2}+N}{N+1}$ (for details see the Supplementary Material). According to LHV theories, $B_{Q}$ is positive. Fig.~\ref{LimitBSV} shows that for sufficiently large $L$, the values of $B_{Q}$ become negative. The mean number of photons in the SV state is $2\sinh^2 \Gamma$. The value of $B_Q$ decreases  with population, $B_Q\approx- e^{2\Gamma}$, and in the macroscopic limit of  $\Gamma \to \infty$,  in the case of $L \to \infty$, we obtain a striking contradiction: $0 \geq \infty \label{paradox}$.

\begin{figure}
\centering 
\includegraphics[height=5cm]{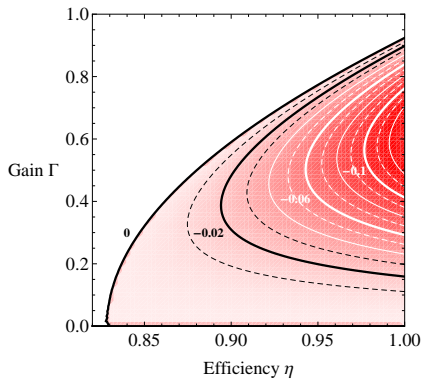}
\caption{Violation of the inequality (\ref{INEQ}) by four mode squeezed vacuum as a function of gain and efficiency, for  two measurement settings.}
\label{LimitBSV2}
\end{figure}

In {\it the case of inefficient detection}, Alice and Bob  measure unequal total photon numbers. This various components $|\psi^{(-)}_{N}\rangle$ of the SV state contribute to the same detection event. We assume that losses in each polarization mode are independent but equal and model them using the Bernoulli distribution with probability of success $\eta$ corresponding to detection efficiency. Probability $p_Q^N(n, m\mid \theta)$ in (\ref{INEQ}) is replaced with a modified one, $P_Q(n, m\mid \theta, \eta)$, which includes all events contributing to a measurement of $n$ photons by Alice and $m$ by Bob with efficiency $\eta$, and the summation over $n$ and $m$ extends to infinity. Fig. \ref{LimitBSV2} displays numerically computed violation of (\ref{INEQ}) as a function of gain and efficiency for the fixed number of settings $L=2$. Violation for the higher gains occurs for larger $L$'s.

{\it Summary and feasibility.} 
We show that for an arbitrary pump power, the {\em four-mode} SV state $\Psi^{(-)}$  involving two propagation and polarization modes, is both an approximation of {\it two copies} of the EPR state and a polarization super-singlet. It has all invariance properties of a two-photon singlet state, although it is a superposition of multi-photon components. We introduce a family of chained Bell inequalities based on the concept of distance, which are  violated by $\Psi^{(-)}$ for all (non-zero) values of squeezing (gain). Our inequalities employ straightforward  local observables: merely  photon numbers at outputs of polarization analyzers, which do not require auxiliary fields, or ancillas; just beam-splitting, no interferometric overlaps. For low pump powers the inequalities do not give results which differ much from the traditional CHSH-like chained inequalities. However, for high powers they are robustly violated because multi-photon emissions do not decrease the contrast of the interference effect which defines the terms of the inequalities (averaged moduli of differences of photon numbers). Note, that standard  correlation functions  $\langle a^{\dagger}_i a_i b^{\dagger}_j b_j \rangle$, which behave as  
$$  \sinh^2{\Gamma}\cosh^2{\Gamma}\cos^2(\theta_{A_i}-\theta_{B_j})+\sinh^4{\Gamma}$$ 
loose their interferometric contrast for increasing $\Gamma$, eventually reaching the value $1/3$, characteristic for thermal fields, see e.g. \cite{LASKOWSKI}. This renders CHSH-like approaches,
based on such correlations, useless. Thus, our chained inequalities are better suited for high gain parametric down-conversion experiments.

The `short'  inequalities (\ref{INEQ}), involving two to four settings at each side, can be useful in quantum information tasks, cryptography and reduction of communication complexity, in device-independent protocols.  They are violated also for final efficiencies, Fig. 5. Note that as the PDC process now produce entangled pairs with fidelity approaching $100\%$, the main distortions in production the SV, which involves multi-pair emissions, are due to losses. Thus our efficiency analysis covers also the imperfections in the generation of SV. 

The inequalities involving large numbers of settings are impractical, but they lead to an "all-versus-nothing" direct GHZ-like refutation of EPR concepts, for states which are close approximations of EPR states and share the basic properties with Bohm's singlets. Thus, the four-mode SV emerges as a versatile state in studies of both quantum information and foundational problems.

The SV states with mean photon number of the order of ten are accessible in laboratories~\cite{Macrobell,PAN}. 
Violations of the presented   Bell inequalities may be soon feasible for small number of settings and for pump intensities in Fig. 5. Experiments could employ the techniques of Ref.~\cite{MOHAMED} and integrated optics setups equipped with superconducting transition-edge sensors (TESs)~\cite{TES1}, which reach photon-counting efficiencies near $100\%$ and have extremely well resolved photon-number peaks, up to around ten photons~\cite{TES2}. Therefore, the efficiency required for the chained Bell inequality violation with the four-mode SV is in principle achievable with state-of-the-art techniques. However, our work is rather a motivation for a new research, than a blue-print for an experiment.

\acknowledgments 

The work is a part of EU project BRISQ2. MS and KR were supported by the EU 7FP Marie Curie Career Integration Grant No. 322150 ``QCAT'', NCN grant No. 2012/04/M/ST2/00789, FNP Homing Plus project No. HOMING PLUS/2012-5/12 and MNiSW co-financed international project No. 2586/7.PR/2012/2. M\.Z is supported by TEAM project of FNP. MW acknowledges NCN Grant No. 2012/05/E/ST2/02352.

\onecolumngrid

\renewcommand{\theequation}{SM.\arabic{equation}}
\setcounter{equation}{0}

\clearpage

\begin{center}
{\Large\bf Supplementary Material:\\ Two Copies of the Einstein-Podolsky-Rosen State of Light\\ Lead to Refutation of EPR Ideas}
\end{center}

\twocolumngrid

\section{Derivation of probability $p_Q^N$ for a singlet state $\psi_N^{(-)}$}

\subsection*{Lossless detection}

We aim at computing the following probability distribution for an arbitrary singlet state $\psi_N^{(-)}$
\begin{align*}
p_Q^N(n,m \mid\theta)=\big\lvert\langle \psi^{(-)}_{N}\vert n, N-n, (N-m)_{H+\theta }, m_{V+\theta}\rangle\big\rvert^2.
\end{align*}
At first, we will express the state $\Ket{(N-m)_{H+\theta }, m_{V+\theta}}$ in terms of the $(H,V)$ basis. The two bases, i.e. $(H,V)$ and $(H+\theta,V+\theta)$ are linked by the following rotation 
\begin{equation}\left(
\begin{array}{c}b_{H+\theta }^{\dagger}\\
b_{V+\theta }^{\dagger}
\end{array}\right) =\left(
\begin{array}{cc}\cos{\theta }&\sin{\theta }\\
                 \- -\sin{\theta }&\cos{\theta }
\end{array}\right) \left( \begin{array}{c}b_{H}^{\dagger}\\b_{V }^{\dagger}
\end{array}\right).
\label{eq:11}
\end{equation}
Using Eq.~(\ref{eq:11}) we obtain
\begin{displaymath}
\begin{split}
&\Ket{(N-m)_{H+\theta }, m_{V+\theta}}=\frac{(b_{H+\theta }^{\dagger})^{N-m}(b_{V+\theta}^{\dagger})^{m}}{\sqrt{(N-m)!m!}}\Ket{0} \\&{}=\frac{1}{\sqrt{(N-m)!m!}}\sum_{p=0}^{N-m}\sum_{q=0}^{m}\binom{N-m}{p}\binom{m}{q}(-1)^{q}\\&\qquad(\sin{\theta })^{N-m-p+q}(\cos{\theta })^{m-q+p}\\&\qquad(b_{H }^{\dagger})^{p+q}(b_{V }^{\dagger})^{N-(p+q)}\Ket{0}
\end{split}
\end{displaymath}
\begin{displaymath}
\begin{split}
&{}=\frac{1}{\sqrt{(N-m)!m!}}\sum_{p=0}^{N-m}\sum_{q=0}^{m}\binom{N-m}{p}\binom{m}{q}(-1)^{q}\\&\qquad(\sin{\theta })^{N-m-p+q}(\cos{\theta })^{m-q+p}\\&\qquad\sqrt{(p+q)!(N-(p+q))!}\Ket{p+q,N-(p+q)}.
\end{split}
\end{displaymath}
We are now ready to compute $p_Q^N(n,m \mid\theta )$. The corresponding probability amplitude is given by

\begin{displaymath}
\begin{split}
&\langle \psi^{(-)}_{N}\vert n, N-n, (N-m)_{H+\theta }, m_{V+\theta}\rangle\\&={}\frac{1}{\sqrt{(N+1)(N-m)!m!}}\sum_{k=0}^{N}(-1)^{k}\sum_{p=0}^{N-m}\sum_{q=0}^{m}\binom{N-m}{p}\binom{m}{q}\\&\qquad(-1)^{q}(\sin{\theta })^{N-m-p+q}(\cos{\theta })^{m-q+p}\\&\qquad\sqrt{(p+q)!(N-(p+q))!}\\&\qquad\langle k, N-k, N-k, k\vert p+q,N-(p+q)\rangle \\&{}=\frac{1}{\sqrt{(N+1)(N-m)!m!}}\sum_{k=0}^{N}(-1)^{k}\sum_{p=0}^{N-m}\sum_{q=0}^{m}\binom{N-m}{p}\binom{m}{q}\\&\qquad(-1)^{q}(\sin{\theta })^{N-m-p+q}(\cos{\theta })^{m-q+p}\\&\qquad\sqrt{(p+q)!(N-(p+q))!}\\&\qquad\delta _{nk}\,\delta _{p+q, N-k}\\
&={}\frac{(-1)^{n}}{\sqrt{(N+1)(N-m)!m!}}\sum_{p=0}^{N-m}\sum_{q=0}^{m}\binom{N-m}{p}\binom{m}{q}\\&\qquad(-1)^{q}(\sin{\theta })^{N-m-p+q}(\cos{\theta })^{m-q+p}\\&\qquad\sqrt{(p+q)!(N-(p+q))!}\,\delta _{p+q, N-n}.
\end{split}
\end{displaymath}
In order to get rid of the Kronecker delta $\delta _{p+q, N-n}$, we notice that $p\in(0, N-m)$ and $0\leq q=N-n-p\geq m-n$. This implies the summation over $q$ from $q=i=\max\{0,m-n\}$ to $j=\min\{N-n,m\}$. Hence, the above probability amplitude simplifies to
\begin{equation}
\begin{split}
&\langle \psi^{(-)}_{N}\vert n, N-n, (N-m)_{H+\theta }, m_{V+\theta}\rangle \\&{}=(-1)^{n}\sqrt{\xi_{nm}^{(N)}}\sum_{q=i}^{q=j}\binom{N-m}{N-n-q}\binom{m}{q}\\&\qquad(-1)^{q}(\sin{\theta })^{2q+n-m}(\cos{\theta })^{N-(2q+n-m)}\\&{}=(-1)^{n}\sqrt{\xi_{nm}^{(N)}}(\cos{\theta })^{N}\sum_{q=i}^{q=j}\binom{N-m}{N-n-q}\binom{m}{q}\\&\qquad(-1)^{q}(\tan{\theta })^{2q+n-m},
\label{amp}
\end{split}
\end{equation}
where $\xi_{nm}^{(N)}=\frac{(N-n)!n!}{(N+1)(N-m)!m!}$. Square of the absolute value of (\ref{amp}) gives the probability $p_Q^N(n,m \mid\theta )$
\begin{align*}
&p_Q^N(n,m \mid\theta )=\xi _{nm}^{(N)}(\cos{\theta })^{2N}\\&\qquad\cdot\Bigg(\sum_{q=i}^{q=j}\binom{N-m}{N-n-q}\binom{m}{q} (-1)^{q}(\tan{\theta })^{2q+n-m}\Bigg)^{2}.
\end{align*}

\subsection*{Imperfect detection}

We assume that the probability distribution describing losses in photon counting detectors is given by the Bernoulli binomial distribution with the success probability (efficiency of detectors) $\eta $. We also assume that losses in each polarization mode are independent but, for simplicity, equal. 

Imperfect detectors will measure $x$ and $y$ photons in modes $a_{H}$ and $b_{V + \theta }$, instead of $n\ge x$ and $m\ge y$, respectively. Therefore, for a $2N$-photon singlet state $\lvert\psi^{(-)}_{N}\rangle$, probability $p_Q^N(x, y\mid n, m, \theta )$ of measuring $x$ and $y$ photons, assuming that before losses there were $n$ and $m$ photons in modes $a_{H}$ and $b_{V + \theta }$, is given by  
\begin{align}
p_Q^N&{}(x, y\mid n, m, \theta, \eta)= \binom{n}{x}\binom{m}{y}\eta ^{x+y}\\
&{}(1-\eta )^{n+m-x-y}p_Q^N(n,m \mid \theta ), \nonumber
\end{align}
where $n+m\leq 2N$. Since the numbers $n$ and $m$ are not known, the probability $p_Q^N(x, y\mid  \theta )$ of detection $x$ and $y$ photons in modes $a_{H}$ and $b_{V + \theta }$ equals
\begin{equation}
P_Q^N(x, y\mid\theta,\eta)= \sum_{n=x}^{N}\sum_{m=y}^{N}p_Q^N(x, y\mid n, m, \theta,\eta)
\label{eq:prob_loss}.
\end{equation}

\section{Violation of chained inequality (8) by a singlet state $\psi_N^{(-)}$} 

\subsection*{Lossless detection}

We rewrite inequality (8) as follows
\begin{align}
\label{eq:13}
&
\sum_{n,m=0}^{N}\lvert m - n\rvert \Lambda (n, m\mid N, \theta )\\&\qquad \geq  
\sum_{n,m=0}^{N}\lvert m - n\rvert p_Q^N(n,m \mid\theta'),\nonumber
\end{align}
where we denote
\begin{displaymath}
\Lambda (n,m\mid N,\theta )= (2L-1)\,p_Q^N(n,m \mid\theta ),
\end{displaymath}
and notice that for $L\to\infty$, $\theta = \tfrac{\pi}{4L} \to0$ and $\theta'= \tfrac{(2L - 1)\pi}{4L}\to\tfrac{\pi }{2}$. 

From now on, the notation $LHS$ and $RHS$ will stand for the left-hand side and the right-hand side of inequality~(\ref{eq:13}). At first, we will consider the RHS. Please notice that
\begin{align}
\label{eq:15}
&\sum_{n,m=0}^{N}\lvert m - n\rvert p_Q^N(n,m \mid\theta' )\\
&{}=\kern-.5em\sum_{n+m<N}^{N}\kern-.5em\lvert m - n\rvert p_Q^N(n,m \mid\theta' )+\kern-.5em\sum_{n+m> N}^{N}\kern-.5em\lvert m - n\rvert p_Q^N(n,m \mid\theta' )\nonumber\\&\qquad{}+\sum_{n+m=N}^{N}\lvert m - n\rvert p_Q^N(n,m \mid\theta' ).\nonumber
\end{align}
In Eq.~(\ref{amp}) we have that $i=q_{min}=\max\{0,m-n\}$ and $j=q_{max}=\min\{N-n,m\}$. For simplicity, we will denote $K(q)=2q+n-m$. Now, we have to consider the following three cases. $1^{\circ }$ If $n+m<N$ then $q_{max}=m$ and $K(q_{max})=n+m<N$. Hence, $K(q)<N$ and $(\cos{\theta'})^{N}(\tan{\theta'})^{K(q)}\rightarrow 0$ if $L\rightarrow \infty$ $(\theta' \rightarrow 0 )$. $2^{\circ }$ Similarly, if $n+m<N$ then $q_{max}=N-n$ and $K(q_{max})=2N-(n+m)<N$. Thus, again $K(q)<N$ and $(\cos{\theta'})^{N}(\tan{\theta'})^{K(q)}\rightarrow 0$. From $1^{\circ }$ and $2^{\circ }$ we conclude that the two first sums in (\ref{eq:15}) vanish for $L\rightarrow \infty$. $3^{\circ }$ If $n+m=N$, we obtain $q_{max}=m=N-n$ and $K(q_{max})=n+m=N$. Since $(\cos{\theta'})^{N}(\tan{\theta'})^{N} \to 1$ and $(\cos{\theta'})^{N}(\tan{\theta'})^{K(q)} \to 0$ for $K(q)<K(q_{max})$, therefore only the last component of the third sum in (\ref{eq:15}) contributes. Since for $n+m=N$ and $q=m$, $\xi_{nm}^{(N)}=\frac{1}{N+1}$ and $\binom{N-m}{N-n-q} \binom{m}{q}=1$, we obtain $p_Q^N(n,m \mid\theta' )\rightarrow \xi_{nm}^{(N)}=\frac{1}{N+1}$, which is grater than $0$. Hence, the expression (\ref{eq:15}) for $L\rightarrow \infty$ takes the following form
\begin{align}
\label{eq:16}
&\sum_{n,m=0}^{N}\lvert m - n\rvert p_Q^N(n,m \mid\theta')\\&\qquad{}=\frac{1}{N+1}\sum_{\substack{n,m=0\\ n+m=N}}^{N}\lvert m - n\rvert\nonumber\\&\qquad{}=\begin{cases} \frac{2}{N+1}\sum_{n=0}^{\frac{N-1}{2}}(N-2n) &\text{for $N$ -- odd,} \\ \frac{2}{N+1}\sum_{n=0}^{\frac{N-2}{2}}(N-2n)
&\text{for $N$ -- even,} \end{cases}\nonumber\\&\qquad{}=\begin{cases} \frac{\frac{1}{2}N^{2}+N+\frac{1}{2}}{(N+1)} &\text{for $N$ -- odd,}\\ \frac{\frac{1}{2}N^{2}+N}{(N+1)} &\text{for $N$ -- even.} \end{cases}\nonumber
\end{align}
Thus, we conclude that for large population
\begin{align}
&RHS \propto N, \quad \textrm{for} \quad L \to \infty. 
\end{align}

\begin{figure}
\centering 
\includegraphics[width=6.5cm]{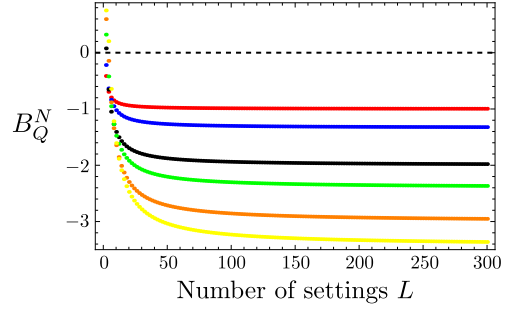}
\caption{Bell parameter $B_{Q}^N$ as a function of number of settings $L$ computed for different photon numbers $2N$: $2N=2$ (red line), $2N=4$ (blue line), $2N=6$ (black line), $2N=8$ (green line), $2N=10$ (orange line), $2N=12$ (yellow line).}
\label{Limit}
\end{figure}

Now, we will show that the function $\Lambda (n,m\mid N,\theta )$ vanishes for $\theta \to 0$ ($L \to \infty$). Inserting  $p_Q^N(n,m \mid\theta)$, $\Lambda$ can be rewritten as follows
\begin{displaymath}
\begin{split}
&\Lambda (n,m\mid N, \theta )= \xi_{nm}^{(N)} \Bigg(\sqrt{2L-1}\cos^N{\theta}\\&\qquad{}\cdot\sum_{q=i}^{q=j}\binom{N-m}{N-n-q}
 \binom{m}{q}(-1)^{q}(\tan{\theta})^{2q+n-m}\Bigg)^{2}.
\end{split}
\end{displaymath}
Please notice that for $m\neq n$ (if $m=n$, $\lvert m - n\rvert=0$) and $q \geq \max\{0,m-n\}$ we have that $K(q)=2q+n-m>0$. Using de L'Hospital rule, we check that $\sqrt{2L-1}\tan{\theta}\to 0$ for $L \to \infty$. From this fact and knowing that $(\cos{\theta})^{N}\to 1$ and $(\tan{\theta})^{l}\to 0$ ($l$ is arbitrary positive number) for $L \to \infty$, we show that  $\sqrt{2L-1}(\cos{\theta})^{N}(\tan{\theta})^{K(q)}\to 0$ for $L \to \infty$. Hence, $\Lambda (n,m\mid N, \theta )\to 0$. From the above considerations we conclude that for an arbitrary $N$
\begin{align}
& LHS\rightarrow 0, \quad \textrm{for} \quad L\to \infty. 
\end{align}

We introduce a Bell parameter in the form of $B_{Q}^N={}\mathrm{LHS} - \mathrm{RHS}$, which according to LHV theories is always positive. Fig.~\ref{Limit} shows $B_{Q}^{N}$ as a function of the number of settings $L$ for various photon numbers $2N$. Independently of $N$, for sufficiently large $L$, $B_{Q}^{N}$ tends to a negative value.

\begin{figure*}[th]
\centering
\raisebox{3.0cm}{(a)}\hskip-0.3cm
\includegraphics[width=5.7cm]{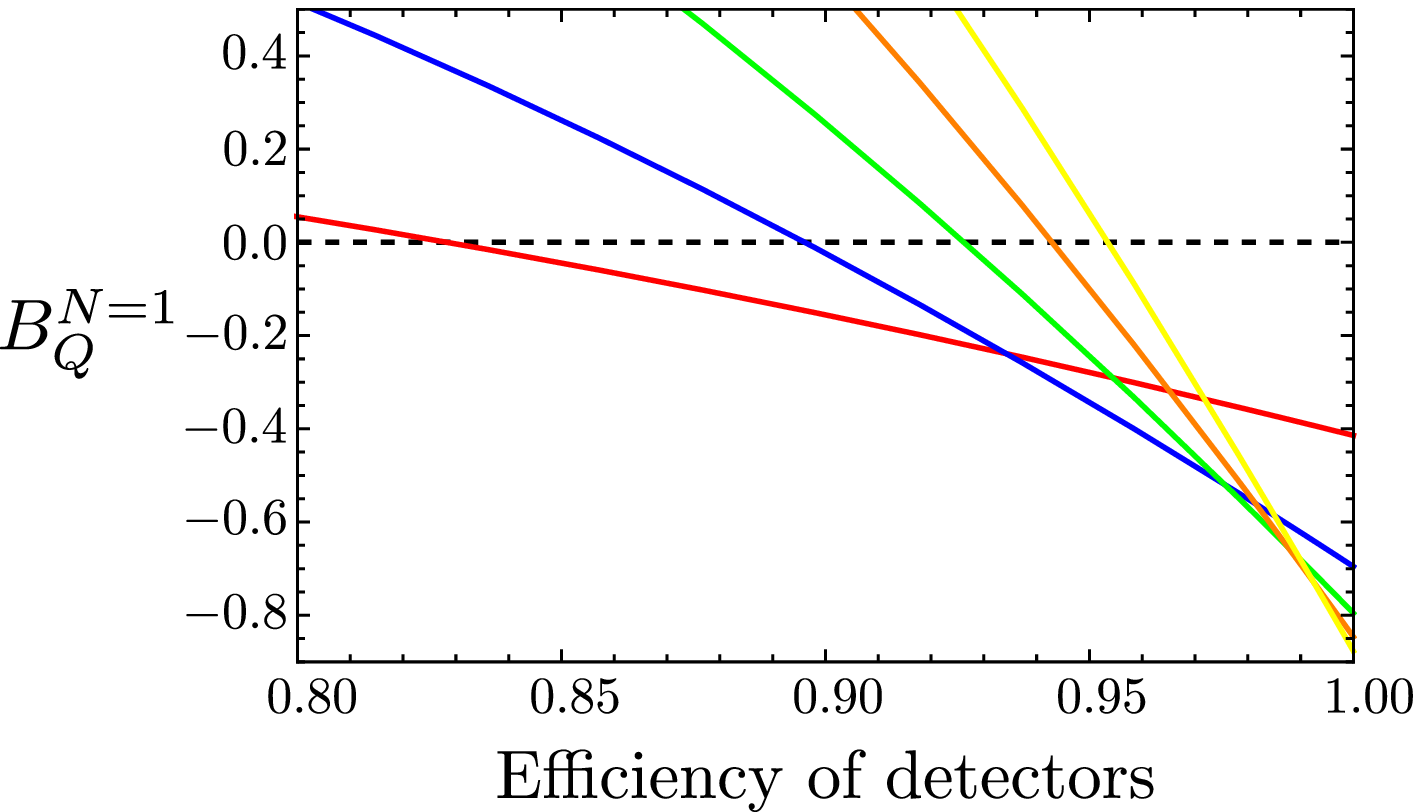}
\raisebox{3.0cm}{(b)}\hskip-0.3cm
\includegraphics[width=5.7cm]{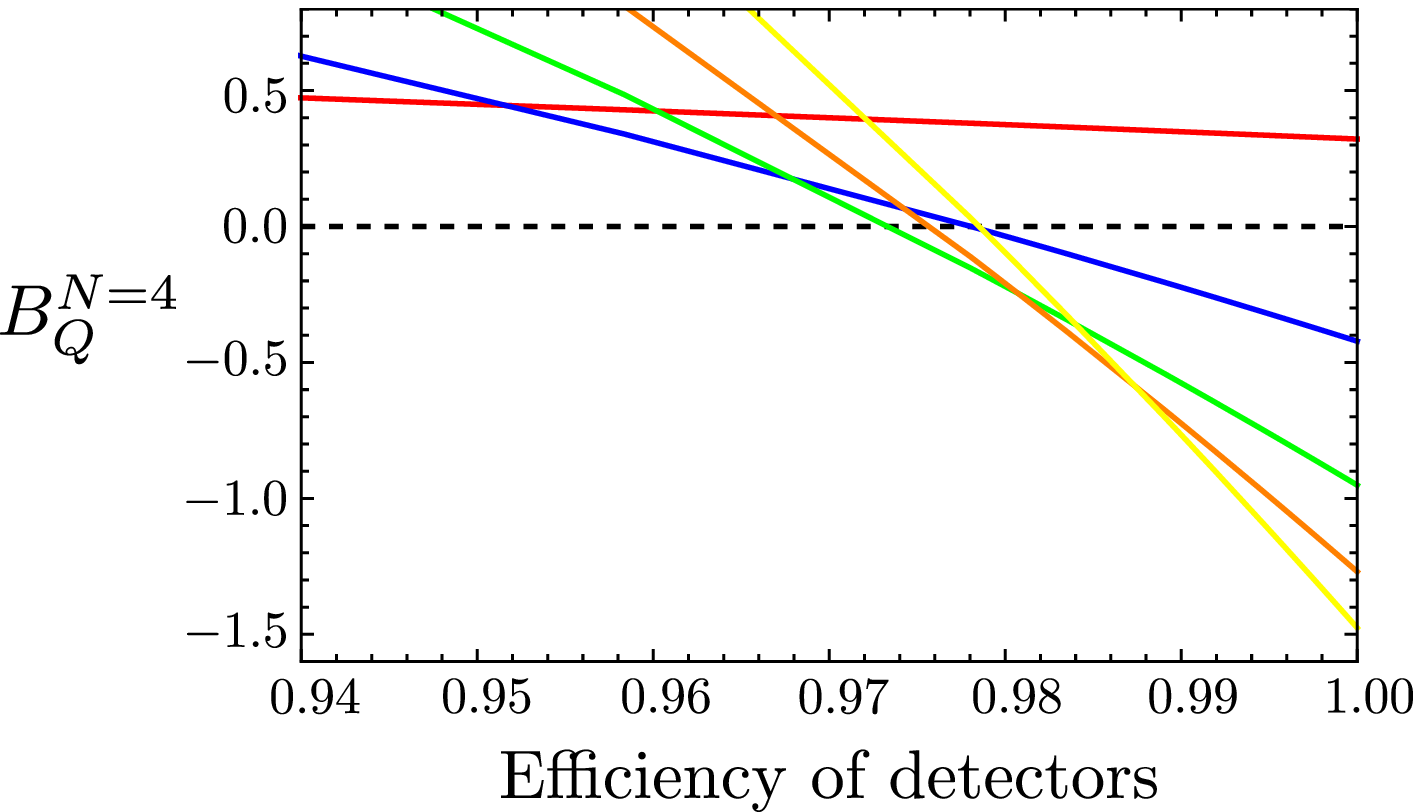}
\raisebox{3.0cm}{(c)}\hskip-0.3cm
\includegraphics[width=5.7cm]{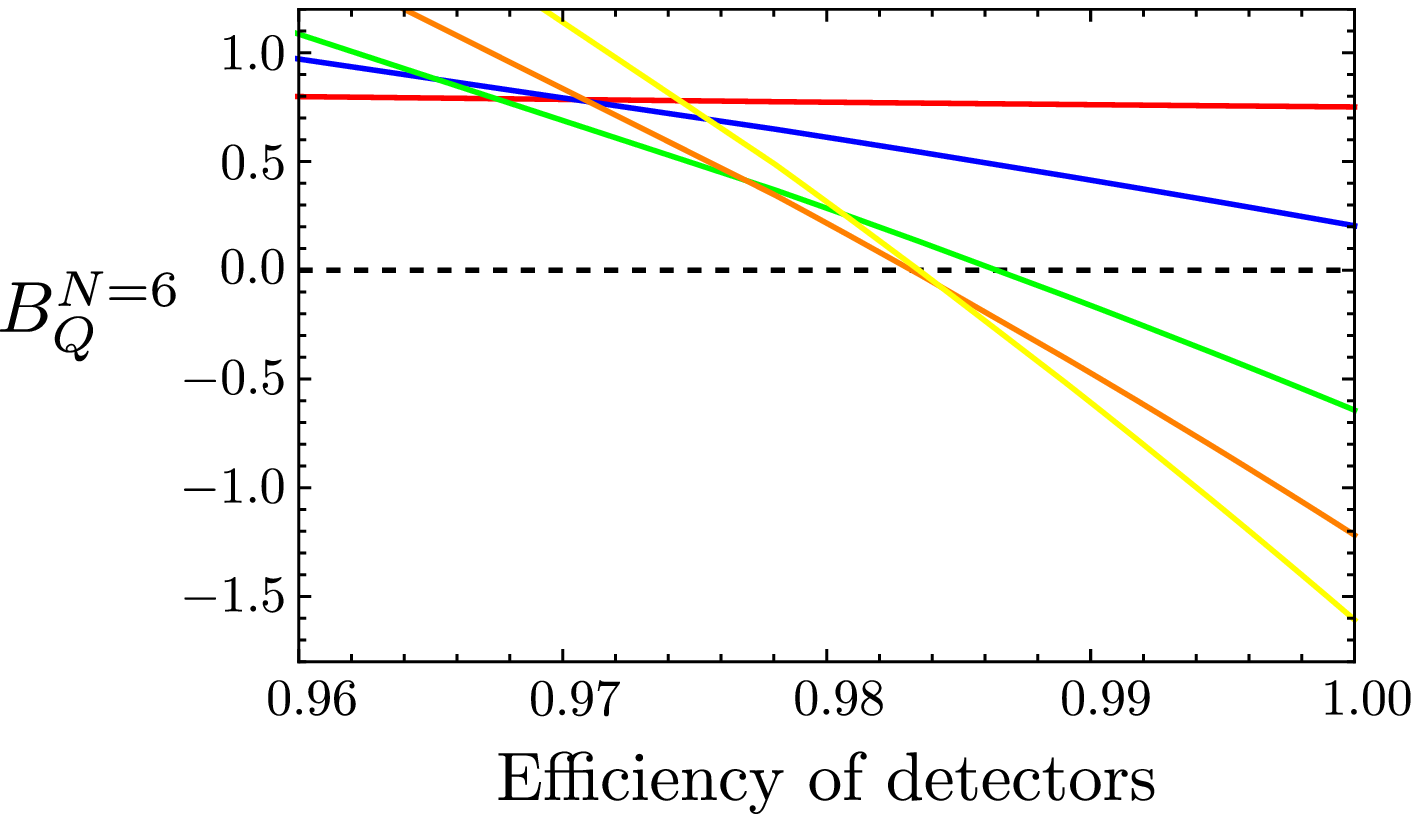}
\caption{Bell parameter $B_{Q}^N$ as a function of efficiency of detectors $\eta $ evaluated for different number of settings $L$: $L=2$ (red line), $L=4$ (blue line), $L=6$ (green line), $L=8$ (orange line), $L=10$ (yellow line) for (a) $2N=2$, (b) $2N=8$ (c) $2N=12$.}
\label{fig:L2N2}
\end{figure*}

\subsection*{Imperfect detection}

We have investigated how fragile is the violation of  $B_Q^N$ in case of a non-unit detection efficiency $\eta$. In order to compute dependence of the Bell parameter on $\eta$, we replace the probability $p_Q^N(n,m \mid\theta)$ with 
$P_Q^N(x,y \mid\theta,\eta)$ given in (\ref{eq:prob_loss}). Inequality (8) is modified as follows
\begin{align}
&
\sum_{x,y=0}^{N}\lvert x-y\rvert \Lambda (x,y \mid N, \theta, \eta )\\&\qquad \geq  
\sum_{x,y=0}^{N}\lvert x-y\rvert P_Q^N(x,y \mid\theta', \eta),\nonumber
\end{align}
where we denote
\begin{displaymath}
\Lambda (x,y\mid N,\theta, \eta )= (2L-1)\,P_Q^N(x,y \mid\theta, \eta ),
\end{displaymath}

Fig.~\ref{fig:L2N2}a displays numerical results obtained for the simplest case $2N=2$ and for various number of settings.  As expected, violation of the local bound is possible only above the usual minimal value of efficiency of detectors $\eta >\frac{2}{1+\sqrt{2}}\approx 83\%$. The violation takes its maximal value $-0.4$ in the limit of perfect detection, $\eta = 1$.  The violation persists for all $L$'s but is most pronounced for the highest (considered) number of settings $L=10$ and reaches $-0.8$. For larger photon numbers, in order to observe any violation, the minimal efficiency as well as the minimal number of settings increase, see Fig.~\ref{fig:L2N2}bc. For $2N=8$ one must have $L\ge 4$ to have a violation. For $2N=12$ the threshold is $L \ge 6$. Interestingly, the minimal efficiency required for violation can be smaller for larger number of settings, compare the green and yellow curves in Fig.~\ref{fig:L2N2}c.

\section{Violation of chained inequality (8) by a squeezed vacuum state (SV) in case of lossless detection} 

We rewrite inequality (8) for a squeezed vacuum state (BSV) (1) in a similar way as we did in the previous paragraph 
\begin{align}
\label{eq:13BSV}
&\sum_{N=0}^{\infty} \lambda_N^2
\sum_{n,m=0}^{N}\lvert m - n\rvert \Lambda (n, m\mid N, \theta )\\&\qquad \geq  
\sum_{N=0}^{\infty} \lambda_N^2 
\sum_{n,m=0}^{N}\lvert m - n\rvert p_Q^N(n,m \mid\theta').\nonumber
\end{align}

Using Eq.~(\ref{eq:16}) and the fact that $\lambda _{N}^{2}=\cosh^{-4}(g)(N+1)\tanh^{2N}(\Gamma )$, we compute the  $RHS$ in the limit of $L \to \infty$ to be
\begin{displaymath}
RHS = \frac{\sinh^{3}(2\Gamma )}{\sinh(4\Gamma )}>0.
\end{displaymath}

Since the $LHS$ for $L \to \infty$ tends to zero independently of $N$ (it was shown in the previous paragraph), we conclude that the Bell parameter defined as $B_Q=LHS-RHS$ of (\ref{eq:13BSV}) takes negative values $B_Q \to -e^{2\Gamma}$.

\section{Violations of chained inequalities (8) by a squeezed vacuum state (SV) in case of imperfect detection}

In the case of imperfect detection, $\eta<1$,  probability of detecting $x$ and $y$ photons by Alice and Bob, respectively, reads
\begin{equation}
P_Q(x, y\mid\theta, \eta) = \sum_{N=0}^{\infty}\lvert\lambda _{N}\rvert^{2}P_Q^N(x, y\mid\theta, \eta),
\end{equation} 
where $\lvert\lambda _{N}\rvert^{2}$ is the probabilistic weight of $\lvert\psi^{(-)}_{N}\rangle$ in the squeezed vacuum state, and $P_Q^N(x, y\mid\theta, \eta)$  were discussed in the previous sections, see (\ref{eq:prob_loss}) . 

However, as we are not able to sum above expression to infinity, we introduce a cut-off parameter $N_{max}$, and define
\begin{equation}
P^{approx}_Q(x, y\mid\theta, \eta, N_{max}) = \sum_{N=0}^{N_{max}}\lvert\lambda _{N}\rvert^{2}P_Q^N(x, y \mid\theta, \eta), \label{PR}
\end{equation}
where 
$N_{max}$ is
such that for a given amplification gain $\Gamma$ we nearly fulfill the normalization condition
\begin{equation}
\sum_{N=0}^{N_{max}}\lvert\lambda _{N}\rvert^{2}\geq 0.99.
\end{equation}

We have carefully checked whether we took large enough values of $N_{max}$ for our numerical computations, so that the results presented in the main text do not change significantly for larger values.

We take the probabilities given in the previous section (\ref{PR}) and directly insert it into inequality (8)
\begin{align}
\label{INEQQ}
{}&\sum_{x,y=0}^{N_{max}}\lvert x - y\rvert (2L-1) P^{approx}_Q(x, y\mid \theta, \eta, N_{max})\\  \geq&{}\sum_{x,y=0}^{N_{max}}\lvert x - y\rvert P^{approx}_Q(x, y\mid \theta', \eta, N_{max}).\nonumber
\end{align}

For the gains $\Gamma<1$, we have checked that $N_{max}=10$ gives a good approximation for the infinite sum. 

The most important numerical results for our discussion,  concerning violation of the inequality (\ref{INEQQ}), are shown in Fig.\ 5 of the main text.
 
\end{document}